# Geodesic Distance Field-based Curved Layer Volume Decomposition for Multi-Axis Support-free Printing

Yamin Li, Dong He, Xiangyu Wang, Kai Tang*

Department of Mechanical and Aerospace Engineering, Hong Kong University of Science and Technology, Clear Water Bay, Kowloon, Hong Kong,

*Corresponding author.

E-mail: ylifm@connect.ust.hk (Yamin Li), dong.he@connect.ust.hk (Dong He), xwangbn@connect.ust.hk (Xiangyu Wang), mektang@ust.hk (Kai Tang)

**Abstract:** This paper presents a new curved layer volume decomposition method for multi-axis support-free printing of freeform solid parts. Given a solid model to be printed that is represented as a tetrahedral mesh, we first establish a geodesic distance field embedded on the mesh, whose value at any vertex is the geodesic distance to the base of the model. Next, the model is naturally decomposed into curved layers by interpolating a number of iso-geodesic distance surfaces (IGDSs). These IGDSs morph from bottom-up in an intrinsic and smooth way owing to the nature of geodesics, which will be used as the curved printing layers that are friendly to multi-axis printing. In addition, to cater to the collision-free requirement and to improve the printing efficiency, we also propose a printing sequence optimization algorithm for determining the printing order of the IGDSs, which helps reduce the air-move path length. Ample experiments in both computer simulation and physical printing are performed, and the experimental results confirm the advantages of our method.

## 1. Introduction

Fused deposition mode lling (FDM) is one of the most popular additive manufacturing technologies, which lays layers by mechanically extruding molten thermoplastic material onto a substrate [1]. Traditionally, most FDM systems are of the 2.5-axis configuration, in which the part model is first sliced into parallel layers with a uniform thickness, and then the material is deposited layer upon layer from bottom-up along a fixed direction (+Z). Referring to Figure 1 (a), under 2.5-axis configuration, the part is decomposed by parallel planes ($S_1, S_2, \ldots, S_i \ldots$). For an arbitrary point $P$ on the boundary curve $B_{i+1}$ of surface $S_{i+1}$, the normal direction $\boldsymbol{n}$ to the plane might not go through the previous plane $S_i$. But we can find a point $P'$ on the boundary curve $B_i$ of surface $S_i$ which minimizes the distance $|\overrightarrow{P'P}|$. The *overhang angle* $\theta$ at $P$ then can be defined as the angle between vector $\overrightarrow{P'P}$ and the normal direction $\boldsymbol{n}$, which identifies the extent of the overhang of plane $S_{i+1}$ to the previous one $S_i$. When the angle $\theta$ is larger than a threshold (e.g., 45°), support structures will be required to avoid the falling of material.

Though simple and hence easy to implement, 2.5-axis configuration has some obvious drawbacks. First, support structures are required when printing overhang features, which will result in the waste of much material and time, and the surface of the part could be easily damaged when removing the support structures manually. Additionally, the staircase effect as induced by parallel slicing often leads to poor surface quality. Earlier efforts have been made to reduce support volume and improve surface quality under the 2.5 axis configuration by optimizing certain printing parameters such as build direction and/or layer thickness [2,3]. However, the improvements in these endeavours are limited due to the nature of the 2.5 axis configuration.

In recent years, multi-axis printing systems have been attracting more and more attention [4-14], which enables the nozzle to change its orientation continuously with respect to the workpiece during the printing process, making it possible to print parts without supports, as well as significantly mitigate the staircase effect. In terms of volume decomposition strategies for multi-axis printing, they can be roughly categorized into the 3+2 axis type and the pure multi-axis type. The methods of 3+2 axis type typically slice the part with a number of unparallel planes and fabricate it along a finite number of fixed directions [9-11]. Refer to Figure 1 (b), the overhang angle of unparallel planes can be reduced compared to that of parallel planes, thereby reducing the support volume. However, this strategy faces difficulty in dealing with freeform parts with complex features, thus lacking generality and flexibility.

As 3+2 axis type methods do not fully utilize the flexibility of the multi-axis system, they are only suitable for prismatic or tree-structured parts. To achieve support-free printing of freeform parts with complex features, curved layer decomposition methods based on the pure multi-axis configuration are becoming a potential solution. As shown in Figure 1 (c), under the paradigm of curved layer decomposition, although the layer thickness is no longer uniform, by continuously adjusting the nozzle orientation when printing the curved surface $S_{i+1}$, the material can always be deposited on the previous curved surface $S_i$, thus in theory completely eliminating the need of any support. Dai et al. [13] proposed a curved layer decomposition method for multi-axis volume printing, in which the part is first represented as a voxel model, then a scalar field called the accumulation field which identifies the printing sequence is established by using the convex-fronts, and finally curved layers are fitted according to the accumulation field and the printing paths are planned on them. Their method is general and capable of printing complex features. However, its decomposition accuracy is sensitive to the voxel density, which requires a large amount of memory and calculation time. Recently, Xu et al. [14] developed a curved layer volume decomposition method for multi-axis printing of freeform parts. In their method, as inspired by the intrinsic characteristics of geodesics, they generated iso-geodesic distance contours on the part surface and used them as the boundary loops of the curved layers, and the slicing surfaces are constructed by filling these loops into surfaces. Though general and algebraically elegant, their method suffers from the fact that, because the slicing surfaces are defined by filling holes of boundary loops, they are susceptible to intersecting each other when the slicing geodesic interval is small (so to achieve a better printing quality) or the geometry of the part is complex with many valleys and peaks.

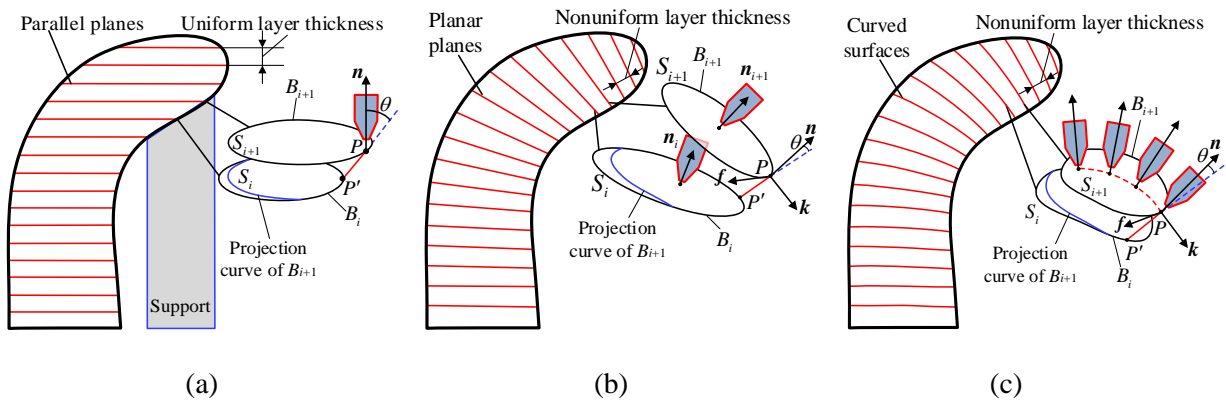

**Figure 1** Illustration of the volume decomposition methods: (a) parallel planar layer decomposition of 2.5 axis printing, support is usually needed; (b) unparallel planar layer decomposition of 3+2 axis printing, support is reduced; (c) curved layer decomposition of pure multi-axis printing, theoretically support is eliminated.

This paper is a continuation of the previous work [14]. Instead of using only the iso-geodesic contours on the part's surface as the boundaries of the curved layers, we compute the geodesic distance field directly inside the 3D volume of the part, which then provides a natural volume decomposition. Referring to Figure 2, for a three-dimensional manifold (a solid), from an initial point or a base of the part, we can define locally parallel geodesics that will fill the entire manifold. The *iso-geodesic distance surfaces* (IGDSs) of this 3D field can be naturally used to decompose the whole part. Because IGDSs are always perpendicular to the geodesics, the overhang angle at the boundary of any IGDS will be significantly reduced.

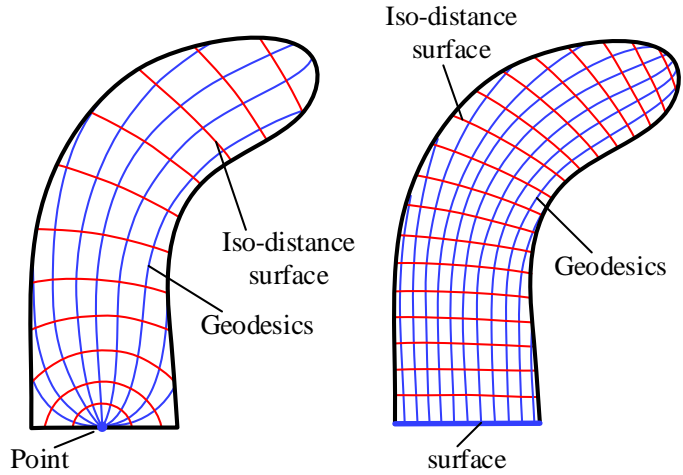

**Figure 2** Geodesics from a point or base inside a three-dimensional manifold and their iso-geodesic distance surfaces.

In terms of the calculation of the geodesic distance field, Crane et al. [15] proposed an efficient and robust method for computing geodesics on a Riemannian manifold. They noticed that the geodesic distance between any pair of points on a Riemannian manifold can be recovered by constructing the heat diffusion field from either point of them because the gradient of the heat field is parallel to that of geodesics. Their method mainly includes three steps: first, the temperature scalar field of the given domain can be obtained by solving the heat flow equation $\dot{u} = \Delta u$ discretely for a fixed time $t$; then the gradient vector field $X$ can be calculated by $X = -\nabla u/|\nabla u|$; finally the geodesic distance field can be determined by solving the Poisson equation $\Delta \phi = \nabla \cdot X$. Figure 3 shows the calculation of geodesics on a triangular mesh.

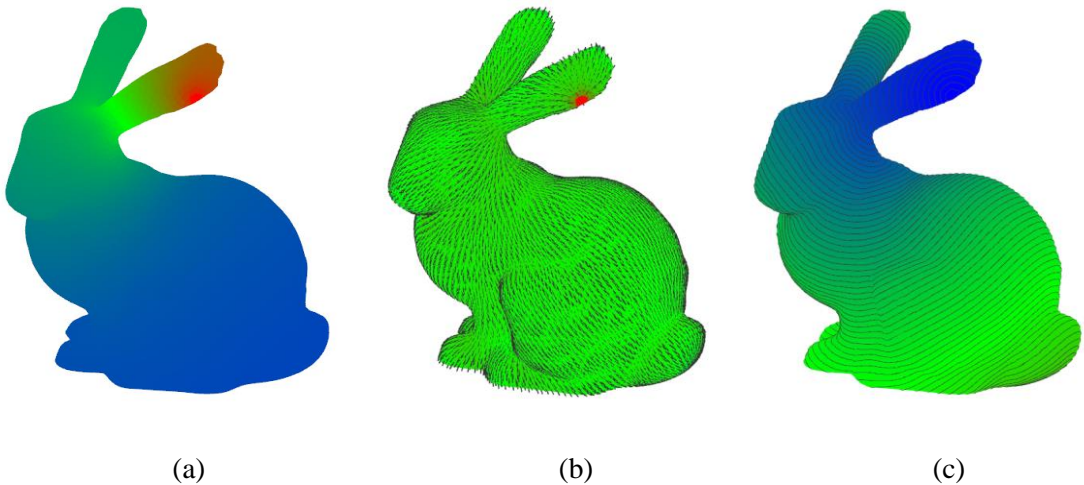

(a)   (b)   (c)

**Figure 3** Illustration of Crane's heat method: (a) the temperature field; (b) the gradient vector field; (c) the iso-geodesic distance contours.

We extend Crane's heat method to the tetrahedral mesh to calculate the geodesic distance field in a solid and then slice the part with a number of interpolated IGDSs. As shown in Figure 4, for a solid to be printed, its geodesic distance field can be calculated first by using Crane's heat method, where the base of the part is generally set as the heat source. Next, the IGDSs are defined by means of interpolation based on this geodesic distance field. These IGDSs are then used as natural slice surfaces on which printing paths are planned. To cater to the requirement of collision-free and the improvement of printing efficiency, we also propose a printing sequence optimization method that can effectively reduce the air-move path length while upholding the collision-free condition. The major advantages of our volume decomposition method and printing sequencing method are the following:

(1) The intersection problem of adjacent slicing surfaces as suffered in the previous work [14] is now effectively eradicated.
(2) Unlike geodesics defined only on the boundary surface of the part, our geodesic distance field defined in the 3D volume of the part offers to be an intrinsic and better representation of the part's geometry; consequently, the IGDSs morphing from bottom-up form a natural partial ordering for multi-axis printing.
(3) The method is robust, efficient, and easy to implement, and its computational complexity is relatively low.

The rest of the paper is organized as follows, in Section 2, we introduce the calculation of the geodesic distance field inside the 3D volume of a solid represented as a tetrahedral mesh, as well as the generation of IGDSs which decompose the whole part. Then Section 3 gives the details of the printing sequence optimization method and the printing path planning method on the IGDSs. In Section 4, to validate the advantages of our method, we report the results of both computer simulation and physical printing experiments of the proposed method on several representative freeform parts. Finally, Section 5 concludes the paper.

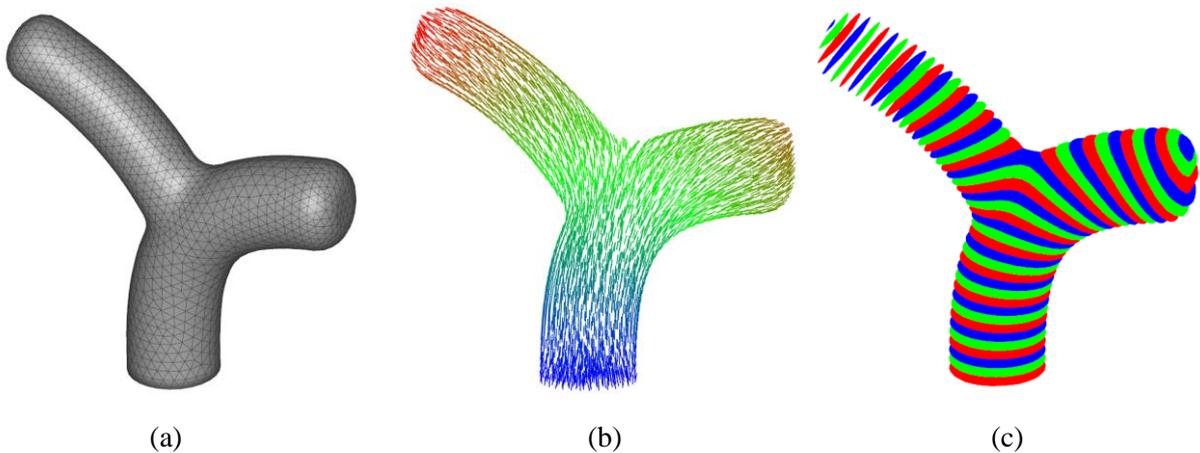

(a)  (b)  (c)

**Figure 4** Illustration of our decomposition method: (a) tetrahedral mesh of the model; (b) geodesic distance field; (c) iso-geodesic distance surfaces (IGDSs).

## 2. Volume decomposition based on geodesic distance field

In this section, we introduce the curved layer volume decomposition method for freeform parts based on the 3D geodesic distance field. Specifically, in Section 2.1, we give the details of computing the 3D geodesic distance field on a tetrahedral mesh. Then, in Section 2.2, we present the algorithm for calculating the IGDSs.

**2.1 Calculation of the geodesic distance field**

We use the Crane's heat method on a tetrahedral mesh to calculate the geodesic distance field of a solid. The first step is to establish the temperature scalar field by solving the heat diffusion equation $\dot{u} = \Delta u$ discretely on the tetrahedral mesh. The discrete form of the heat diffusion equation can be written as

$$(\mathbf{I} - t\Delta)\boldsymbol{u}_t = \boldsymbol{u}_0 \tag{1}$$

where $\mathbf{I}$ is the identity matrix, $\boldsymbol{u}_0$ is the initial temperature field vector, and $\boldsymbol{u}_t$ is the temperature field vector at moment $t$. In order to solve Eq.(1) on a tetrahedral mesh, we need to define the discrete Laplacian operator $\Delta L$ for vertex $i$, which can be expressed as

$$\Delta L = \frac{1}{V_i} \sum_{j \in N(i)} w_{ij}(u_i - u_j) \tag{2}$$

where $V_i$ is one fourth the volume of all tetrahedrons incident on vertex $i$, $N(i)$ is the set of vertices immediately adjacent to vertex $i$, and $w_{ij}$ is a scalar weight assigned to edge $(i, j)$. Liao et al. [16] derived a weight expression for vertex $i$ on a tetrahedral mesh, which is the sum of the components of its $m$ adjacent tetrahedrons

$$w_{ij} = \frac{1}{6} \sum_{k=1}^{m} l_k \cot(\theta_k) \tag{3}$$

Figure 5 illustrates the definitions of $l_k$ and $\theta_k$ in Eq. (3). For the $k$th tetrahedron adjacent to edge $(i, j)$, $l_k$ is the length of the opposite edge $(p, q)$ to which edge $(i, j)$ is against, while $\theta_k$ is the dihedral angle of edge $(p, q)$. Laplacian operator can be expressed via a matrix for all the $n$ vertices as

$$\mathbf{L} = \mathbf{V}^{-1}\mathbf{L}_c \tag{4}$$

where $\mathbf{V} \in \mathbb{R}^{n \times n}$ is a diagonal matrix containing the vertex volumes, and $\mathbf{L}_c \in \mathbb{R}^{n \times n}$ is the Laplacian matrix whose value $L_{ij}$ can be given by

$$L_{ij} = \begin{cases} -\sum_{v_k \in N(i)} w_{ik}, & i = j \\ w_{ij}, & j \in N(i) \\ 0, & otherwise \end{cases} \tag{5}$$

Then Eq. (1) can be expressed as

$$\left(\mathbf{I} - t\mathbf{V}^{-1}\mathbf{L}_c\right)\boldsymbol{u}_t = \boldsymbol{u}_0 \tag{6}$$

and the temperature field $\boldsymbol{u}_t$ at moment $t$ can be obtained by solving Eq. (6), while the appropriate time step $t$ can be calculated by $t = h^2$, where $h$ is the average length of edges [15].

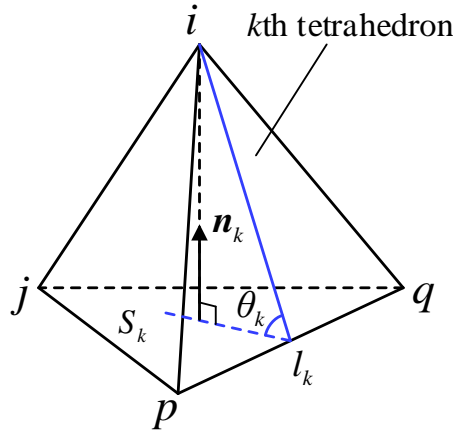

**Figure 5** Illustration of the discrete volumetric Laplacian operator

After the calculation of the temperature scalar field, the temperature gradient $\boldsymbol{g}_k$ inside any tetrahedron ($i$, $j$, $p$, $q$) can be calculated by [16]

$$\boldsymbol{g}_k = \begin{bmatrix} (\boldsymbol{v}_i - \boldsymbol{v}_q) \\ (\boldsymbol{v}_j - \boldsymbol{v}_q) \\ (\boldsymbol{v}_p - \boldsymbol{v}_q) \end{bmatrix}^{-1} \begin{bmatrix} 1 & 0 & 0 & -1 \\ 0 & 1 & 0 & -1 \\ 0 & 0 & 1 & -1 \end{bmatrix} \begin{bmatrix} u_i \\ u_j \\ u_p \\ u_q \end{bmatrix} \tag{7}$$

where $\boldsymbol{v}_i$, $\boldsymbol{v}_j$, $\boldsymbol{v}_p$, and $\boldsymbol{v}_q$ are the vertices' coordinates. The gradient $\boldsymbol{g}_k$ should be normalized ($\boldsymbol{g}_k = \boldsymbol{g}_k / |\boldsymbol{g}_k|$) as the gradient vector of geodesic distance field. The integrated divergence of the gradient field associated with vertex $i$ can be written as

$$\left(\text{Div}\boldsymbol{g}\right)(\boldsymbol{v}_i) = \sum_{k \in N(i)} -\frac{S_k \boldsymbol{n}_k \cdot \boldsymbol{g}_k}{3} \tag{8}$$

as shown in Figure 5, where $S_k$ and $\boldsymbol{n}_k$ are the area and the normal vector of the opposite triangular face $(j, p, q)$ to vertex $i$, respectively, and $N(i)$ is the set of tetrahedrons immediately adjacent to vertex $i$. Finally, the geodesic distance field $\phi$ for all the vertices can be obtained by solving the following equation

$$\mathbf{L}_c \phi = \boldsymbol{b} \tag{9}$$

where $\boldsymbol{b}$ is the divergence field vector of the gradient field.

## 2.2 Calculation of the iso-geodesic distance surfaces

Once the geodesic distance field of the tetrahedral mesh is obtained, we can generate via interpolation a set of IGDSs to decompose the whole part. The interpolation point for a given geodesic distance $\phi$ locates at an edge $(i, j)$ that satisfies the condition $\phi_i < \phi < \phi_j$, where $\phi_i$ and $\phi_j$ are the geodesic distances at the two vertices of edge $(i, j)$; then, the coordinate of the linear interpolation point $\boldsymbol{v}$ can be written as

$$\boldsymbol{v} = \left(\left|\phi_i - \phi\right|\boldsymbol{v}_j + \left|\phi_j - \phi\right|\boldsymbol{v}_i\right) / \left|\phi_i - \phi_j\right| \tag{10}$$

where $\boldsymbol{v}_i$ and $\boldsymbol{v}_j$ are the coordinates of vertex $i$ and $j$.

For a given $\phi$, in order to construct the corresponding IGDS, we first need to connect the interpolation points within any involved tetrahedron into triangles. A tetrahedron may contain either three interpolation points (Figure 6 (a)) or four interpolation points (Figure 6 (b)). For the first case, we can naturally define a triangular face. While for the second, two triangular faces can be defined by adding an edge along one of the two diagonals of the quadrilateral. We call these triangular faces the interpolation faces (of the given $\phi$). Next, starting from any such interpolation face, due to the continuity of the geodesic distance field, we "grow" to another interpolation face that shares one edge with the current one, and then the next, and the next, until all the interpolation faces are traversed, which together form a connected IGDS of $\phi$. In case there are more than one connected IGDS, we move to a remaining untraversed interpolation face, start the growing process again to generate another connected IGDS, and then another, until all the connected IGDSs are found and constructed.

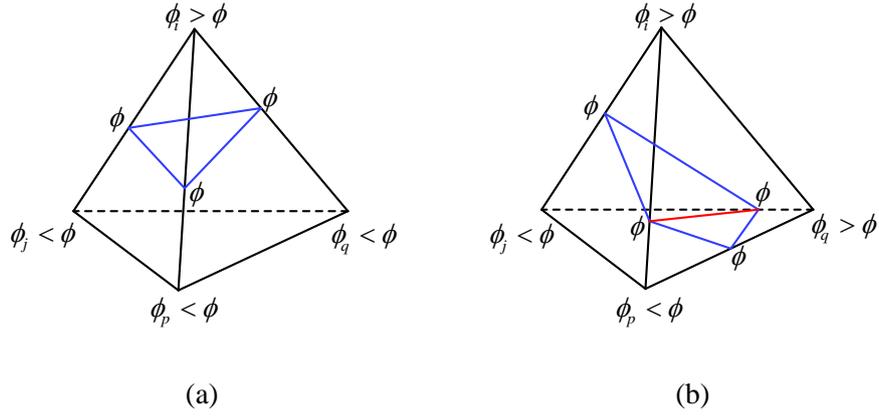

(a) (b)

**Figure 6** Illustration of interpolation points on the edges of a tetrahedron and their triangular faces: (a) three interpolation points; (b) four interpolation points.

Take the Y model shown in Figure 7 as an example. The tetrahedral mesh of the model contains 4014 vertices and 17260 tetrahedrons. We set the vertices at the bottom face as the heat source, and the corresponding geodesic distance field is shown in Figure 7 (a), with a maximum geodesic distance of 44.68 mm, and the whole part is sliced into 44 layers by setting the geodesic interval to be 1 mm, as shown in Figure 7 (b). Each layer in general is bounded by two neighbouring IGDSs. As shown in Figure 7 (c), the mesh quality of any initial IGDS obtained by direct interpolation is not high, which may not be suitable for planning a printing path on it. As an improvement, we apply the well-known isotropic remeshing method [17] and the Laplacian smoothing method [18] in tandem to enhance the mesh quality of the initial IGDS, as demonstrated in Figure 7 (d).

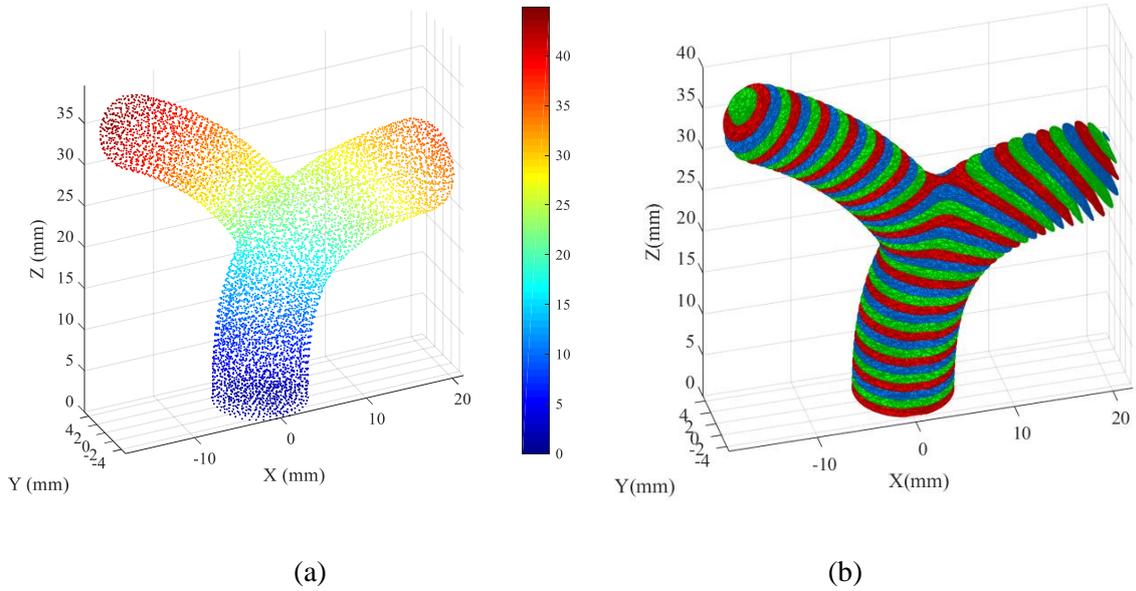

(a) (b)

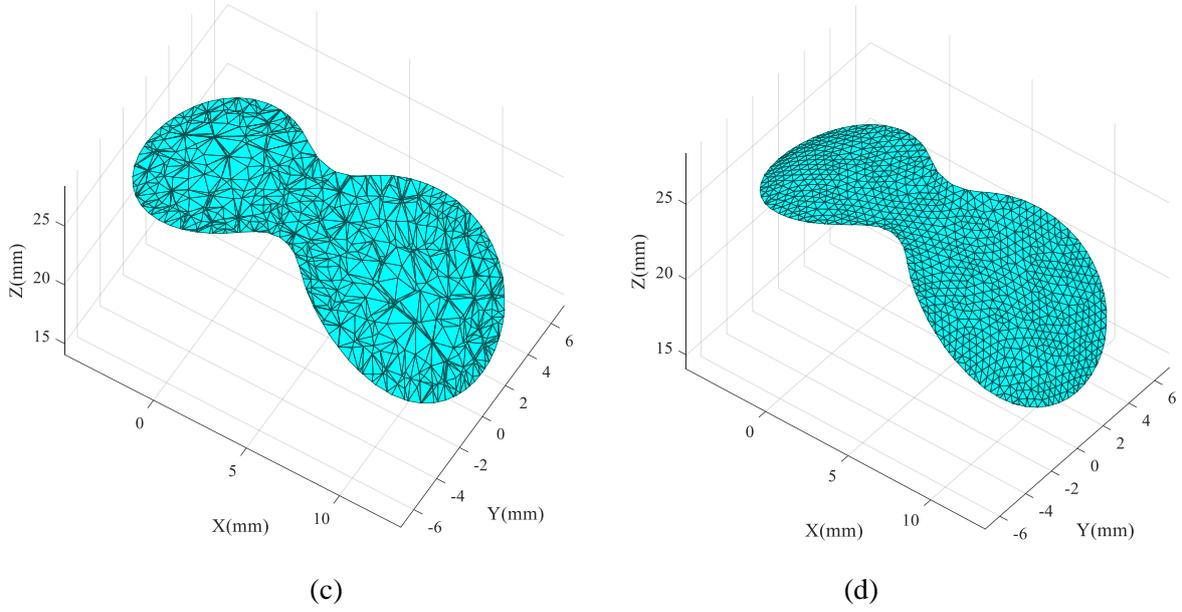

| (c) | (d) |

**Figure 7** Calculation of the IGDSs: (a) geodesic distance field; (b) triangular meshes of the IGDSs obtained by interpolation; (c) triangular mesh of the IGDS at the 24$^{th}$ layer; (d) triangular mesh after isotropic remeshing and Laplacian smoothing.

Because IGDSs are always perpendicular to the geodesics which are locally parallel to each other, the overhang angles of the IGDSs are effectively reduced as a whole, making it possible to print a part without any support. (Note that, as a convention, we enforce that the nozzle axis be aligned with the normal vector of the IGDS at the point of print.) As illustrated in Figure 8, the average overhang angle $\bar{\alpha}$ and the overhang ratio $r$ of an IGDS are defined in Eq. (11), where the averaging is in respect to the boundary curve of the IGDS:

$$\begin{cases} \bar{\alpha} = \dfrac{\int \alpha(s)ds}{S_o} = \dfrac{\int \alpha(s)ds}{\int f(s)ds} \\ r = \dfrac{S_o}{S_c} = \dfrac{\int f(s)ds}{\int ds} \end{cases}, \quad f(s) = \begin{cases} 1, & \alpha(s) > 0 \\ 0, & \alpha(s) = 0 \end{cases} \tag{11}$$

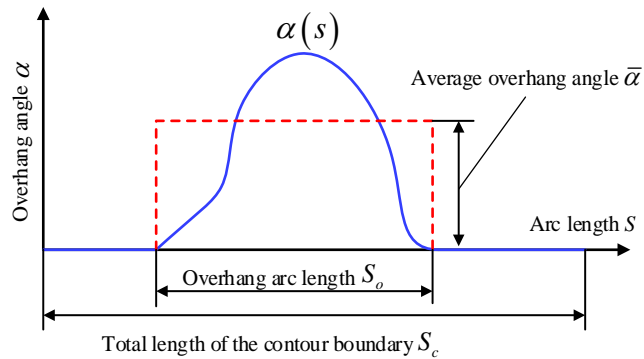

**Figure 8**  Definition of average overhang angle and overhang ratio

Figure 9 (a) shows the distribution of overhang angle at different IGDSs of the Y model (given in Figure 7) when the geodesic distance interval is 1 mm, where it can be seen that large overhang angles occur near the juncture of the three branches. Figure 9 (b) displays graph of the average overhang angle of the IGDSs when the geodesic distance interval is 1 mm, 1.5 mm and 2 mm respectively. It can be seen that the three graphs have a similar pattern, which indicates that the overhang angle is an intrinsic value which is majorly determined by the distribution of the geodesics in the solid. A similar behaviour can be found for the overhang ratio, as shown in Figure 9 (c).

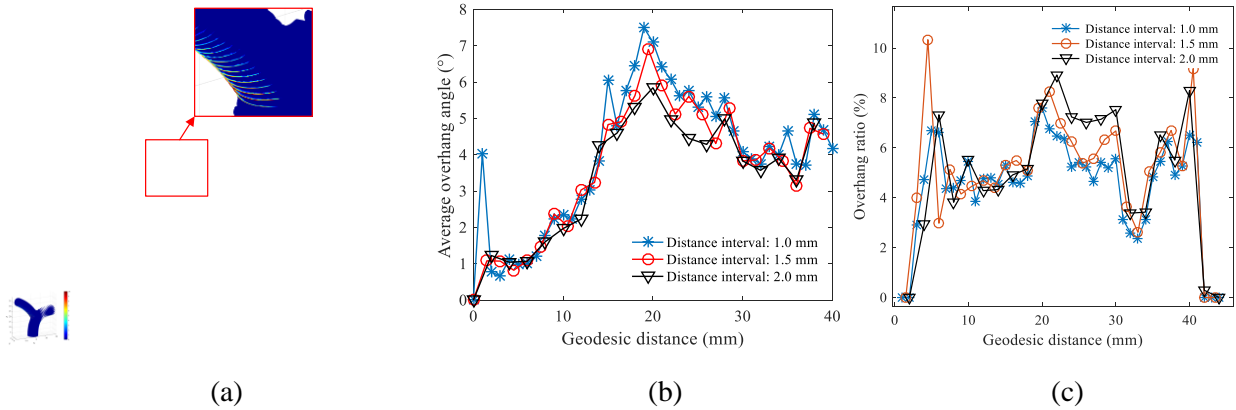

(a)  (b)  (c)

**Figure 9**  Illustration of the overhang angle of the IGDSs: (a) distribution of overhang angle at different layers; (b) average overhang angle at different layers; (c) overhang ratio at different layers.

Although adjacent IGDSs share the same geodesic distance interval with each other and have smaller overhang angles compared to parallel planes, the layer thickness between any two adjacent IGDSs is not uniform due to the intrinsic nonuniform distribution of geodesics. For the Y model, Figure 10 (a) shows the distribution of layer thickness deviation at different layers when the geodesic distance interval is 1mm, with a maximum percentage deviation of 35%, and Figure 10 (b) shows the statistical results. Similar to the overhang angle, larger thickness deviations occur at the boundary curves of the IGDSs.

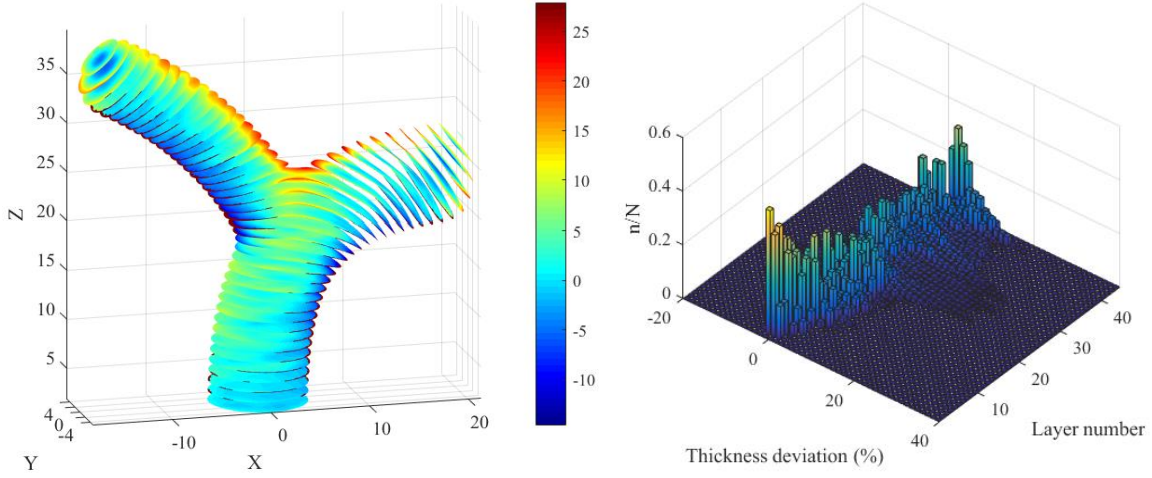

**Figure 10** Illustration of layer thickness of the IGDSs: (a) distribution of layer thickness deviation at different layers; (b) statistical chart of layer thickness deviation at different layers.

After the construction of IGDSs, we define a tree data structure simply called a *skeleton tree* that identifies the topological relationship between them. The whole part is decomposed into a number of layers according to the geodesic distance, with each layer sandwiched between adjacent IGDSs. As aforementioned, for any geodesic distance $\phi$, there could be more than one corresponding IGDS. Notation wise, let $\Psi = \{\phi_1 = d, \phi_2 = 2d, ..., \phi_i = id, ...\}$ be the set of sampling geodesic distances and $S_i = \{S_{i,1}, S_{i,2}, ..., S_{i,j}, ...\}$ be the set of IGDSs corresponding to $\phi_i$. As illustrated in Figure 11, on the skeleton tree, each node represents an IGDS, and every pair of adjacent IGDSs is corresponded by an edge between the two representative nodes on the tree. For any node *a* (IGDS) on the skeleton tree, if node *b* (IGDS) is connected to *a* by an edge on the tree, we call *b* an upper-node of *a* if *b*'s geodesic distance is larger than *a*'s, and a lower-node otherwise. For example, on the tree in Figure 11, node $S_{25,1}$ has two upper-nodes, i.e., $S_{26,1}$ and $S_{26,2}$, and only one lower-node, $S_{24,1}$. The skeleton tree is constructed from the bottom towards the top, and the corresponding algorithm is given in Algorithm 1, wherein function *AreTwoSurfacesAdjacent* ($S_{i,j}$, $S_{i+1,k}$) is used to judge whether any two IGDSs $S_{i,j}$ and $S_{i+1,k}$ are adjacent to each other (i.e., to be connected by an edge). Referring to Figure 12, to judge whether $S_{i,j}$ and $S_{i+1,k}$ are adjacent, we can first randomly select an edge on the triangular mesh (the boundary of the part) that intersects the boundary of $S_{i,j}$; then, from this edge, we trace out a geodesically steepest ascending path on the triangular mesh. Similarly, we can also trace out a geodesically steepest descending path from $S_{i+1,k}$. The two IGDSs are adjacent to each other if at least one of the two paths goes through both.

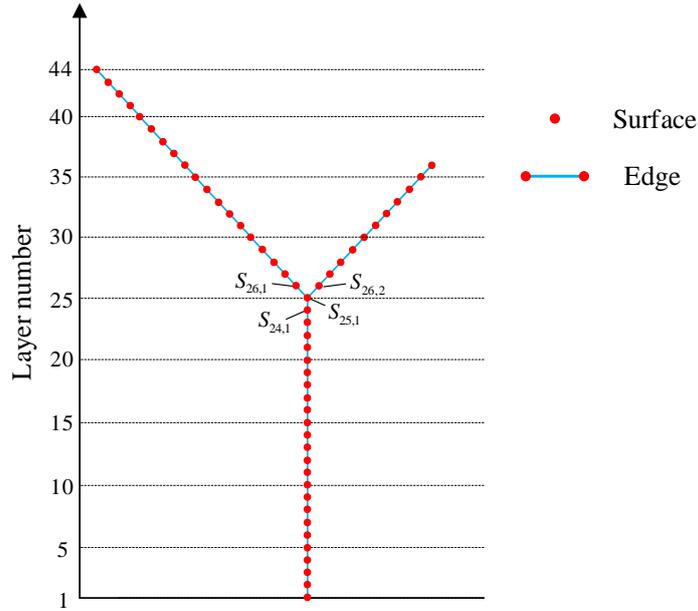

**Figure 11** Generating the skeleton tree of the IGDSs

Algorithm 1 Construction of the skeleton tree

---

**ALGORITHM 1: Construction of the growing tree**

**Input:** All the IGDSs of the decomposed layers, $S_i = \{S_{i,1}, S_{i,2}, ..., S_{i,j}, ...\}$

**Output:** Skeleton tree of the IGDSs

1  integer $n$ = the number of layers
2  **for** $i = 1 : n\text{-}1$
3      **for** each IGDS $S_{i,j}$ of layer $S_i$ **do**
4          **for** each IGDS $S_{i+1,k}$ of layer $S_{i+1}$ **do**
5              **if** *AreTwoSurfacesAdjacent* ($S_{i,j}, S_{i+1,k}$) **then**
6                  Construct an edge between the two contour nodes
7              **end**
8          **end**
9      **end**
10 **end**

---

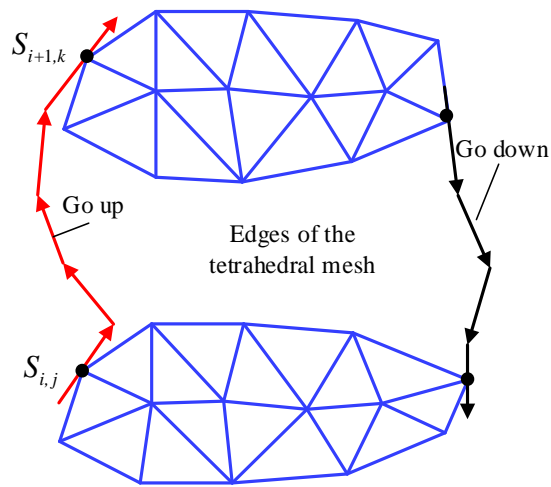

**Figure 12**  Judge whether the two IGDSs are adjacent to each other

## 3. Printing process planning

We have now decomposed the whole part into a number of layers sandwiched between IGDSs and constructed the corresponding skeleton tree which identifies the topological relationship of the IGDSs. In this section, we will first present our printing sequence generation method (Section 3) and then describe how the printing paths are planned on the IGDSs (Section 3.2).

### 3.1 Printing sequence generation

The strict increasing geodesic order of the skeleton tree has already defined a partial order of printing – any node must be printed before its upper-node. However, since printing is a time-continuous process, we must convert this partial ordering into a total ordering of traversal of the nodes, i.e., a single sequence of nodes to print, called a printing sequence. Hereafter we will interchangeably use the terms a "node" and an IGDS. The following criterion must be satisfied for any valid printing sequence:

**Criterion 1**: an IGDS can only be printed if its lower IGDS(s) have already been printed.

Generally, there are two traversal strategies of a skeleton tree, i.e. the layer priority traversal (*LPT*) and the depth priority traversal (*DPT*). Referring to Figure 13, the layer priority traversal strategy traverses the skeleton tree layer by layer from bottom-up, which tends to avoid the collision to the utmost, for that the IGDSs of each layer share the same geodesic distance. On the other hand, the depth priority traversal strategy traverses the skeleton tree along branches in priority, which favours minimizing the air-movement of the nozzle. However, as shown in Figure 14, under DPT, if a branch grows too deep, it may cause collisions when printing other branches.

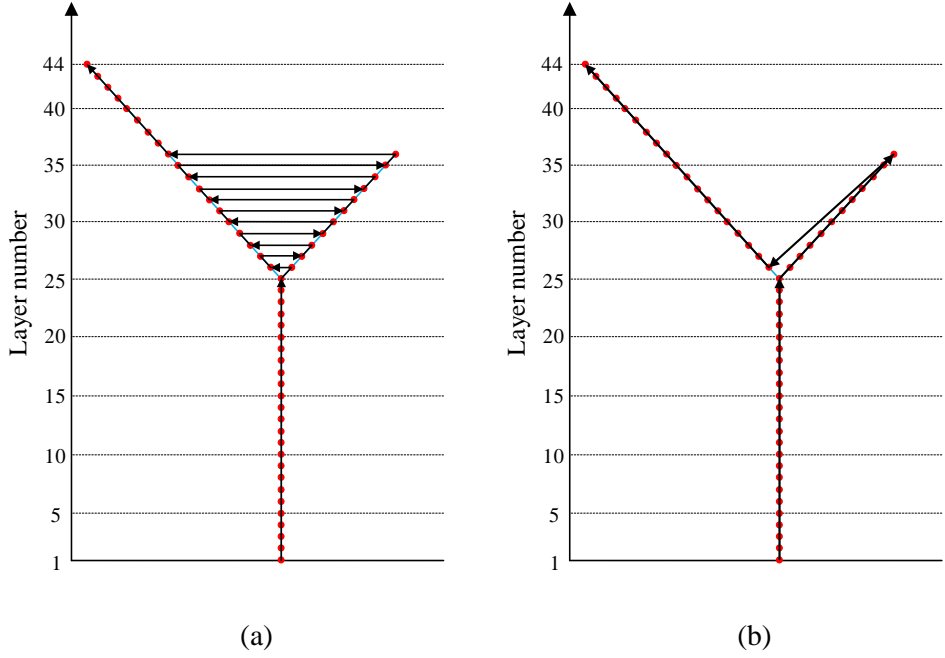

**Figure 13** Traversal strategies of a skeleton tree: (a) layer priority traversal; (b) depth priority traversal

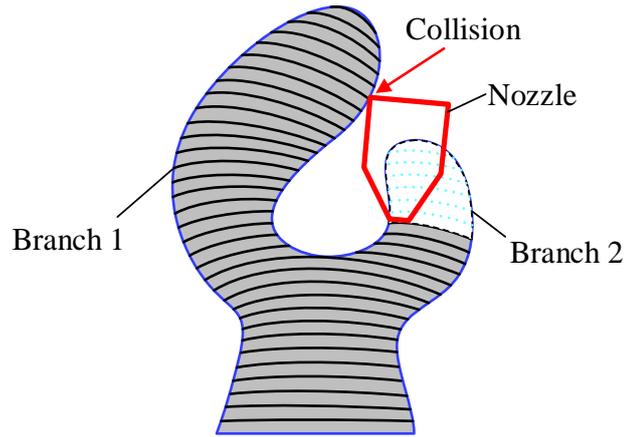

**Figure 14** Illustration of possible collisions during a depth priority traversal

In order to reduce the air-move path length while ensuring no collisions, we propose an optimization method which seeks a compromise between LPT and DPT. First of all, the collision check between the nozzle and the IGDSs must be modelled. In this paper, the shape of the nozzle is simplified by its bounding cone, as shown in Figure 15 (a). Admittedly, this simplification is too conservative, However, because collision check is not the main topic of this paper, we simply choose this simplification to implement our algorithm. When the nozzle cone sweeps along the boundary curve of an IGDS with its orientation coincident with the surface normal direction $n$, the envelope of motion will be a ring-like ruled surface $S(u,v) = P(u) + v k(u)$, where $P(u)$ is an arbitrary point on the boundary curve of the IGDS, and the unit vector $k(u)$ of the generator can be obtained by rotating the normal vector $\boldsymbol{n}$ at $\boldsymbol{P}(u)$ around the tangent vector $\boldsymbol{\tau}(u)$ of the boundary curve with the nozzle angle $\alpha$, as shown in Figure 15 (a). Specifically, to construct the triangular mesh of the ruled surface, we first place a few sample points on the boundary curve of the IGDS, then calculate their generators, and finally connect the generators as triangles. The upper and lower holes of the ring-like ruled surface should be filled to approximate the envelope volume of the cone over the entire IGDS, the lower hole can be filled by the IGDS directly, while the upper hole can be filled by some hole filling algorithms, in this paper, we adopt the advancing front mesh (AFM) technique [19] to patch the hole, which is robust and simple. To determine whether there is a potential collision when printing this IGDS, we only need to check whether there are intersections between other IGDSs and this envelope volume. For each IGDS, we can calculate all the *potential collision surfaces* (PCS) (i.e., other IGDSs which intersect with the envelope volume of this IDGS). Take the part shown in Figure 15 (b) as an example, the potential collision surfaces for surface $S_{4,1}$ will be $S_{5,1}$, $S_{6,1}$, $S_{7,1}$, $S_{7,2}$ and $S_{8,2}$. The detailed procedures for calculating the PCSs of each IGDS are given in Algorithm 2, where function *CollisionCheck* ($S_i$, $S_j$) is used to judge whether there are intersections between mesh surface $S_j$ and the envelope volume of mesh surface $S_i$ – it returns true if an intersection is identified and false otherwise.

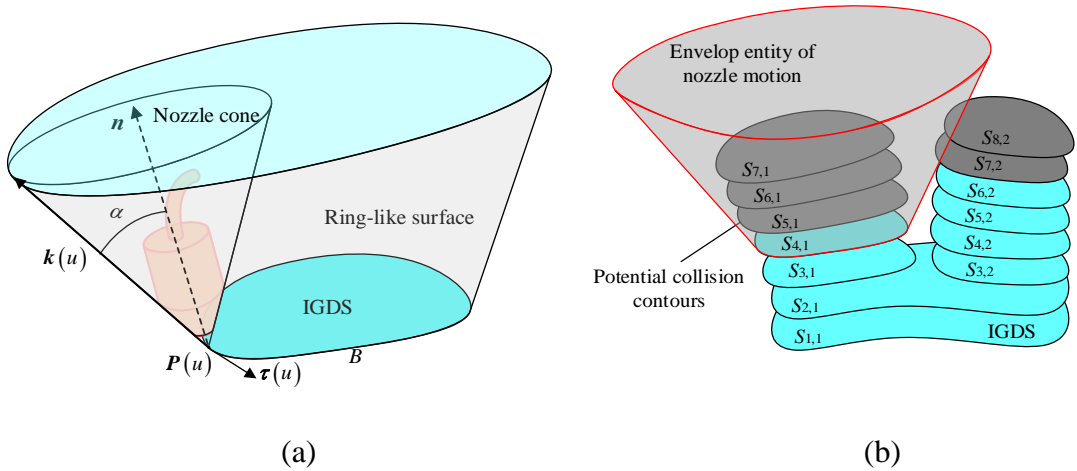

**Figure 15** Illustration of collision check between the nozzle and an IGDS: (a) ring-like surface generated by sweeping the nozzle along the boundary of the IGDS; (b) envelope volume of the nozzle motion

Algorithm 2 Calculation of the potential collision surfaces of each IGDS

|  | **ALGORITHM 2:** Calculation of the PCSs of each IGDS |
|---|---|
|  | **Input:** the IGDSs list $\{S_1, S_2, \ldots, S_i, \ldots\}$ and the nozzle cone angle $\alpha$ |
|  | **Output:** PCSs list for each IGDS |
| 1 | **integer** $k$ = total number of IGDSs |
| 2 | **for** $i = 1: k$ **do** |
| 3 |     **for** $j = 1: k$ **do** |
| 4 |         **if** $i\ !=j$ **then** |
| 5 |         **if** *CollisionCheck* ($S_i$, $S_j$) **then** |
| 6 |             Put $S_j$ into the PCSs list of $S_i$. |
| 7 |         **end** |
| 8 |       **end** |
| 9 |     **end** |
| 10 | **end** |

Facilitated by the PCSs of each IGDS, we propose a greedy traversal (*GT*) algorithm that can generate a collision-free printing sequence with a shorter air-move path length than that of the layer priority traversal. Besides Criterion 1, another criterion must be satisfied during the traversal:

**Criterion 2**: an IGDS can only be printed if the PCSs of the unprinted IGDSs do not include this IGDS.

Specifically, in Algorithm 3 below, function *UpdateNodeCadidates* (*ST*) is used to find all the printable node candidates which satisfy both Criterion 1 and Criterion 2 according to the current skeleton tree *ST*. And function *SelectANode* (*NC*, $N_c$) is used to update the current node $N_c$ from the node candidates list *NC*. It will select the upper-node(s) of $N_c$ in priority. However, if the upper-node(s) are not included in *NC*, the node which is nearest to $N_c$ will be selected. To summarize, we traverse the skeleton tree along the branches in priority unless a potential collision is encountered.

Algorithm 3 Printing sequence optimization algorithm

|   |   |
|---|---|
|   | **ALGORITHM 3:** Printing sequence optimization |
|   | **Input:** The skeleton tree *ST* of the IGDSs |
|   | **Output:** Printing sequence list *PQ* of the IGDSs |
| 1 | Node candidates list *NC* = *UpdateNodeCadidates* (*ST*) |
| 2 | Current node $N_c$ = either node in *NC* |
| 3 | **while** *NC* != ∅ **do** |
| 4 |     Label $N_c$ as printed |
| 5 |     Put $N_c$ into *PQ* |
| 6 |     *NC* = *UpdateNodeCadidates* (*ST*) |
| 7 |     $N_c$ = *SelectANode* (*NC*, $N_c$) |
| 8 | **end** |

**3.2 Printing path planning**

As mentioned in Sec. 2.2, the layer thickness *h* varies at different positions of an IGDS. Additionally, to facilitate path planning, the path step over *l* is set to be a constant during the printing process, which will then require a variable filament feed rate. Refer to Figure 16, the intersection of the extruded filament is assumed to be a rectangle of *h*×*l*; then the following mass conservation equation should be satisfied during the printing process

$$\pi r_m^2 f_m = \mu l h f_p \tag{12}$$

where $r_m$ is the radius of original filament, $f_m$ is the feed rate of the filament, $f_p$ is the feed rate of the nozzle, and $\mu$ is a correction coefficient which is smaller than 1 and can be determined by experiments.

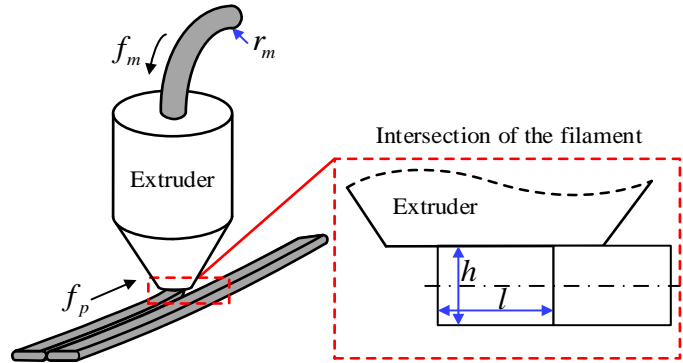

**Figure 16** Illustration of the printing parameters

Similar to the work [14], given an IGDS, its iso-geodesic distance contours can be used as a filling path, which ensures a constant step-over between every two adjacent paths. As shown in Figure 17 (a), also by using Crane's heat method on the triangular mesh of the IGDS, the geodesic distance field with respect to the boundary can be generated, the iso-geodesic distance contours can be computed, and they are then connected as the filling path, as shown in Figure 17 (b) and (c).

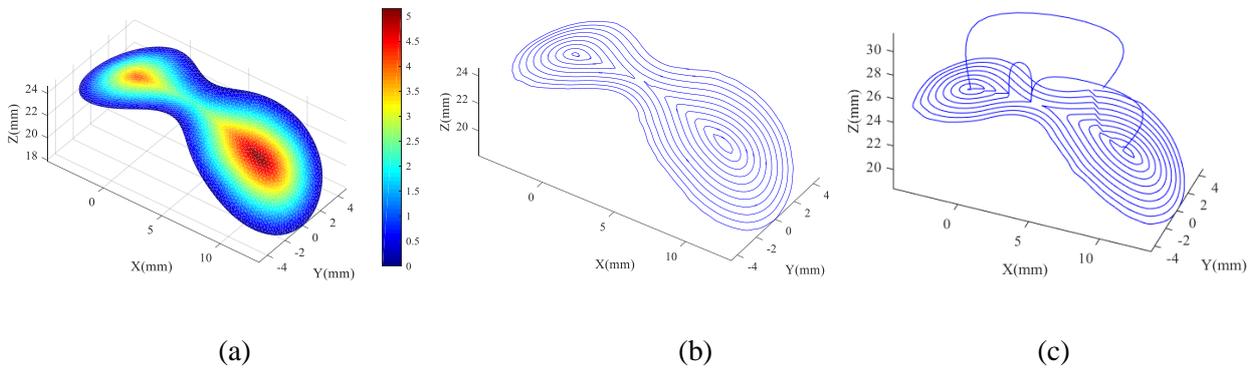

(a)  (b)  (c)

**Figure 17** Filling path based on the geodesic distance field of the contour: (a) geodesic distance field with respect to the boundary contour; (b) iso-geodesic distance contours; (c) connected filling path

## 4. Experiment and discussion

We have implemented the proposed method in C++ and run the program on a laptop with an Intel i7 CPU. In this section, we report the experimental results of our method on four exemplary parts. Three free-form parts with complex features (e.g., overhangs and genus-one structures) are chosen to show the generality of the proposed method. In addition, a tree-structured part with three branches is tested to demonstrate the advantages of the printing sequence optimization method in reducing the air-move path length.

## 4.1 Printing of free-form parts

Three free-form parts (i.e., the Y model, the bunny, and the kitten, shown in Figure 18) are decomposed into IGDS layers by using our method. The tetrahedral meshes of the three parts are generated by the commercial software *HyperMesh*, and the number of tetrahedrons of the Y model, the bunny, and the kitten is 17235, 61579, and 58146, respectively. The geodesic distance interval is set to be 0.6 mm, and the corresponding number of IGDS layers of the three parts is 182, 152 and 153 respectively (see Figure 19). Figure 20 and Figure 21 show the average overhang angle and the overhang ratio distributions of the three parts at different layers, where the average overhang angles of the Y model and the kitten are all smaller than 45°. However, the average overhang angles of the bunny at some layers where sharp overhang features grow are larger than 45°, indicating the possibility of material flow.

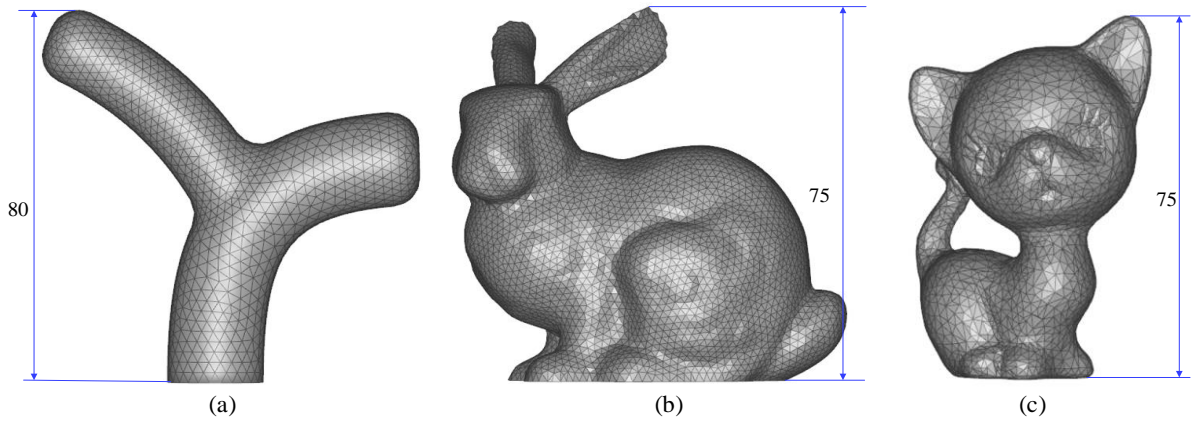

**Figure 18** Three free-form parts to be printed: (a) Y model; (b) bunny; (c) kitten.

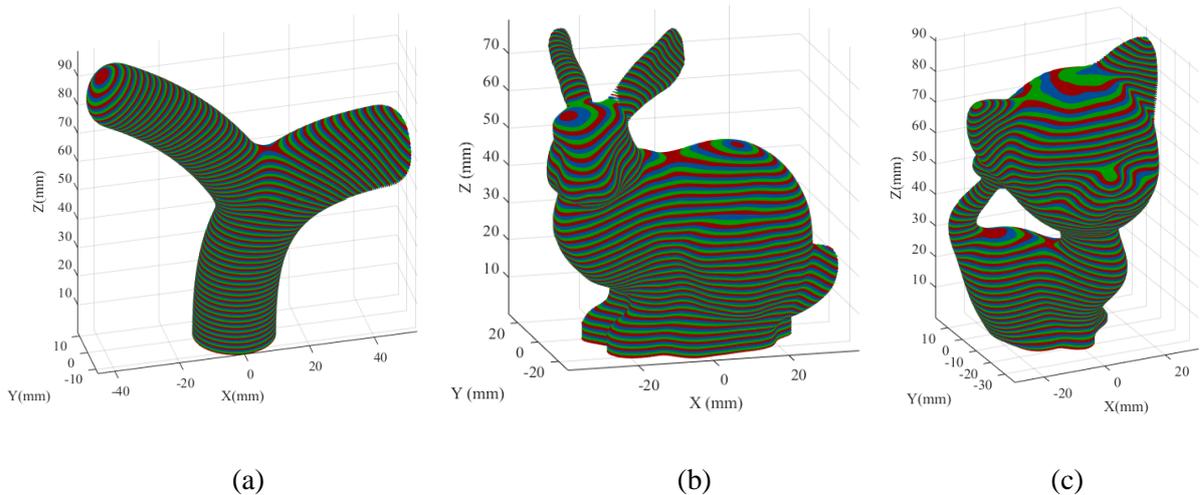

**Figure 19** IGDS layer decomposition of the three parts: (a) Y model, 182 layers; (b) bunny, 152 layers; (c) kitten, 153 layers.

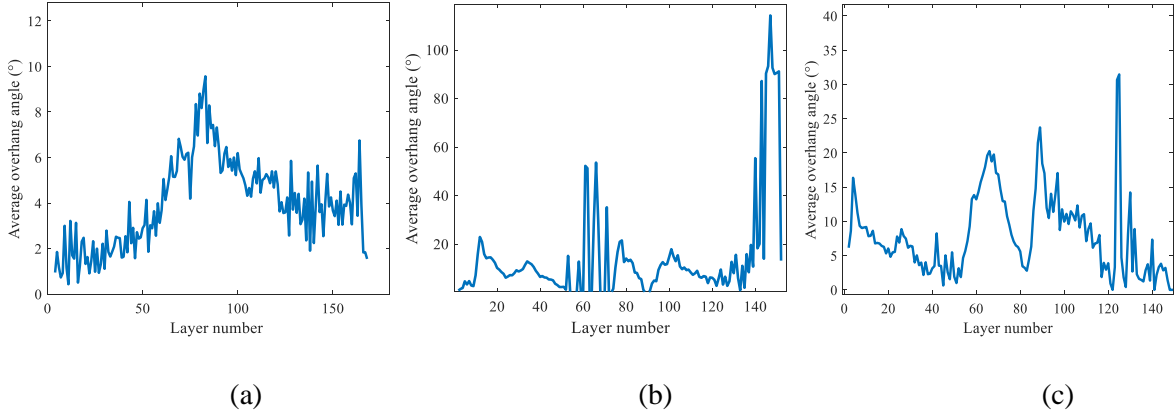

**Figure 20** The average overhang angle distributions of the three parts: (a) Y model; (b) bunny; (c) kitten.

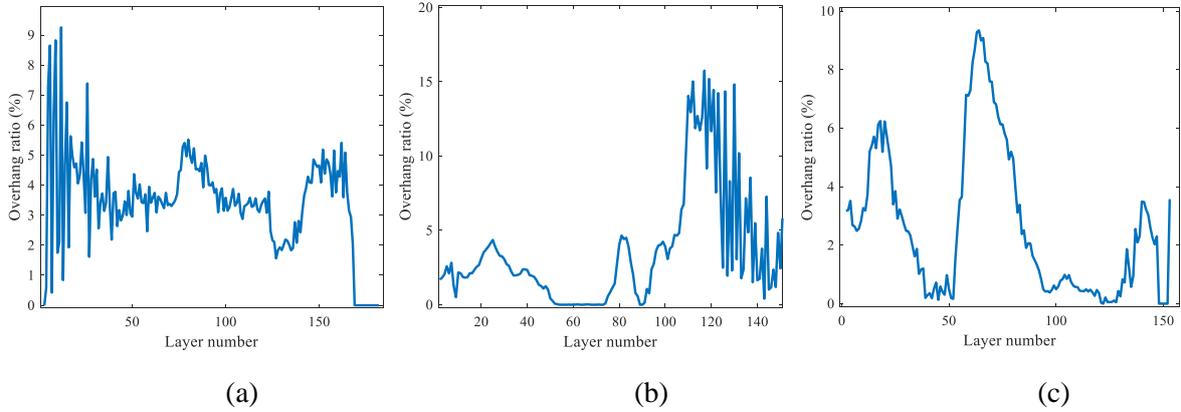

**Figure 21** The overhang ratio distributions of the three parts: (a) Y model; (b) bunny; (c) kitten.

In terms of the computational cost, as our volume decomposition method is a combination of several processes, Table 1 lists the time complexities of these processes and the actual amounts of running time of the three tests. The calculation of the geodesic distance field involves solving a linear system, so the time complexity is only $O(n)$, where $n$ is the number of mesh vertices. The time complexity of the algorithm of IGDSs interpolation is $O(n*m)$, where $n$ and $m$ are the number of layers and the average number of mesh triangles of each layer. The time complexities of the isotropic remeshing and the Laplacian smoothing are at most $O(n^2)$ and $O(n)$, respectively, where $n$ is the number of mesh triangles. It can be found that the isotropic remeshing takes the most running time. We compare the running time of Xu's method [14] with that of our method in Table 2, and it can be seen that our method is much more efficient. Take the bunny as an example: our method decomposes the part into 152 layers within 55 s, while Xu's method takes much more time (693 s) to decompose the same part into only 60 layers.

Table 1 Time complexity of the algorithms and running time of the three tests

| Process | Algorithm | Time complexity | Running time (s) |
|---|---|---|---|
| Tetrahedral mesh generation | *HyperMesh* | / | Y model: 2; Bunny: 3; Kitten: 3 |
| Geodesic distance field generation | Sec. 2.1 | $O(n)$ | Y model: 2; Bunny: 3; Kitten: 3 |
| IGDSs interpolation | Sec. 2.2 | $O(n*m)$ | Y model: 3; Bunny: 5; Kitten: 4 |
| Isotropic remeshing | Ref. [17] | $O(n^2)$ | Y model: 24; Bunny: 45; Kitten: 41 |
| Laplacian smoothing | Ref. [18] | $O(n)$ | Y model: 1; Bunny: 2; Kitten: 2 |

Table 2 Running time comparison with Xu's method [14]

| Method | | Number of layers | Time for layer decomposition (s) |
|---|---|---|---|
| Our method | Bunny | 152 | 55 |
| | Kitten | 153 | 52 |
| Xu's method [14] | Bunny | 60 | 693 |
| | Kitten | 61 | 310 |

Our homemade five-axis multi-axis printing system is shown in Figure 22, which is composed of a multi-axis robot arm (UR5) and a three-axis filament feed system. The robot enables the part fastened on it to reach a desirable posture with respect to the nozzle. The robot and the filament feed rate are synchronously controlled to ensure that Eq.(12) is always satisfied during the printing process. The detailed description of our printing system can be referred to [12, 14].

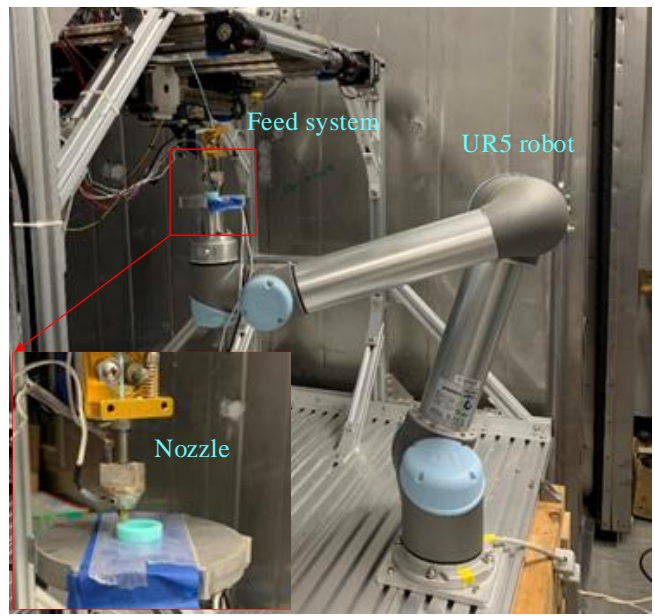

**Figure 22** Multi-axis printing system

The printing path step over is set to be 0.8 mm, and the printing sequence of IGDSs are generated according to the principle of layer priority traversal. Figure 23 shows the printing paths generated by our method, Figure 24 shows the actual printing processes of the three parts, and Table 3 shows the detailed printing parameters. The Y model and the bunny were successfully printed without any supports (see Figure 24 (a) and (b)), and good surface quality was achieved in both finished parts. The IGDSs morph from bottom-up in an intrinsic and smooth way due to the nature of geodesics, which avoid large orientation changes, thereby increasing the printing efficiency.

However, we failed to print the kitten due to the unresolvable local collision between the nozzle and the in-process workpiece, as shown in Figure 25 (a). IGDSs cannot always maintain convex due to the intrinsic topology of the workpiece, which may cause local collisions or gouges when the nozzle angle is large. This limitation is especially obvious for genus-n parts. The kitten is a genus-one part, it can be found in Figure 25 (c) that the geodesics first split at the place where multi-branch begins and merge again at the top of the genus one structure. Due to the misalignment of the geodesics from the two branches, large curvature IGDSs will be generated at places where they begin to merge. One solution to this problem is using a slender nozzle which can avoid local collisions or gouges, as shown in Figure 25 (b); another possible solution is properly adapting the geodesic field, letting the geodesics from different branches become more aligned when they merge, so that the corresponding IGDSs will be more gentle smooth, and consequently avoid collisions. These will be our future research topics.

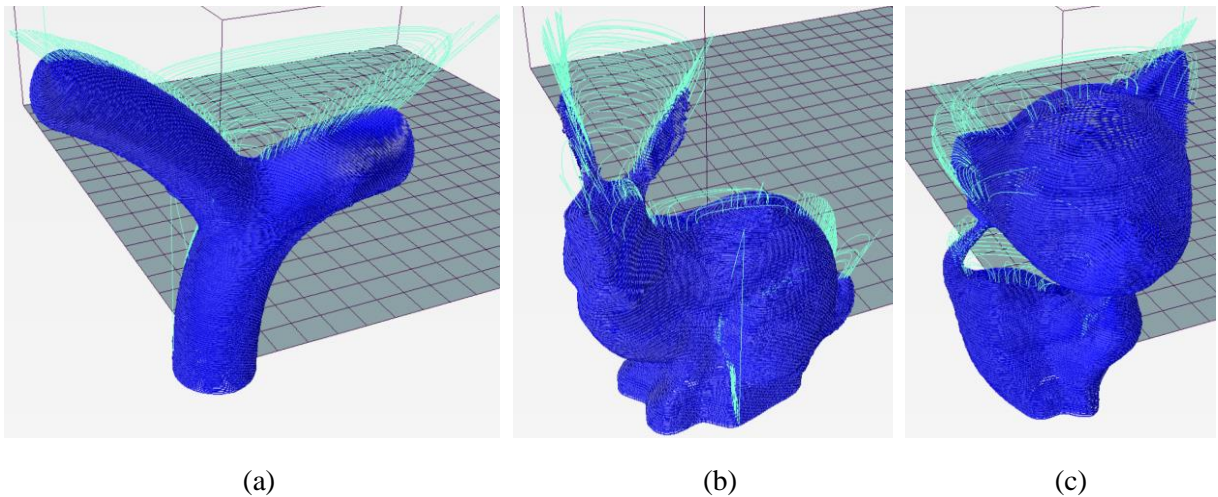

(a) (b) (c)

**Figure 23** The generated printing paths of the three parts and computer simulation: (a) Y model; (b) bunny; (c) kitten.

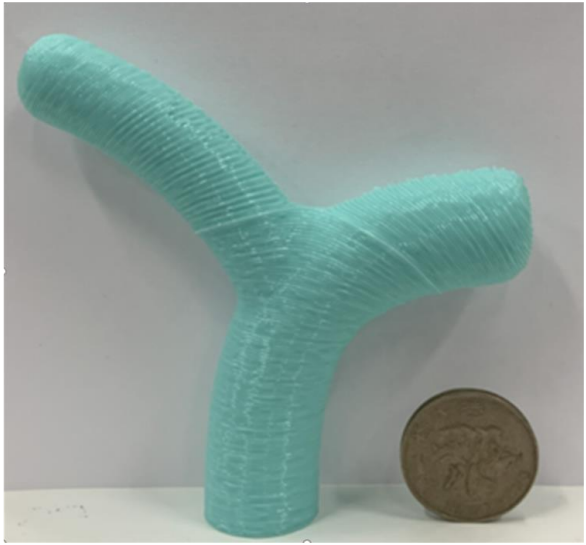
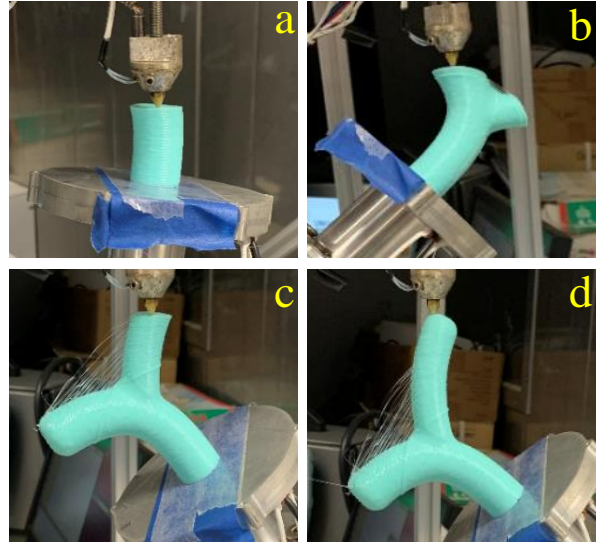

(a)

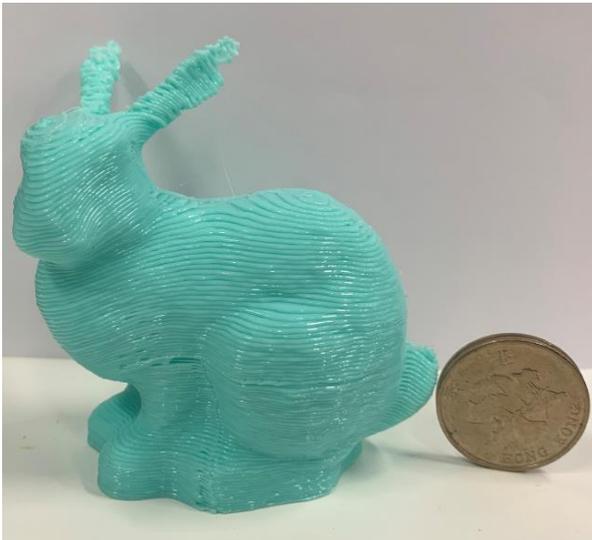
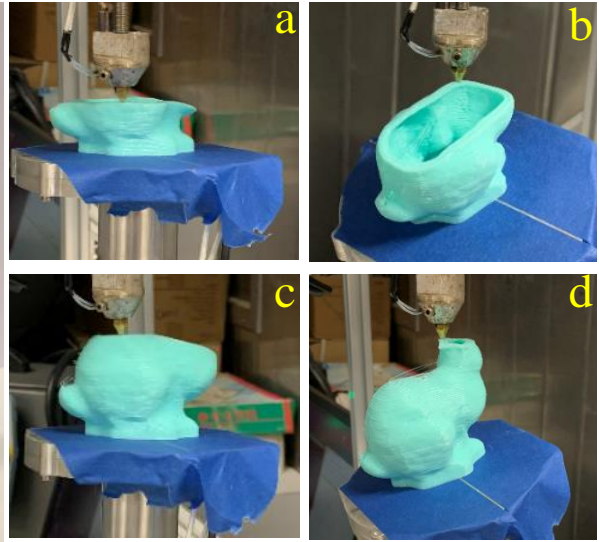

(b)

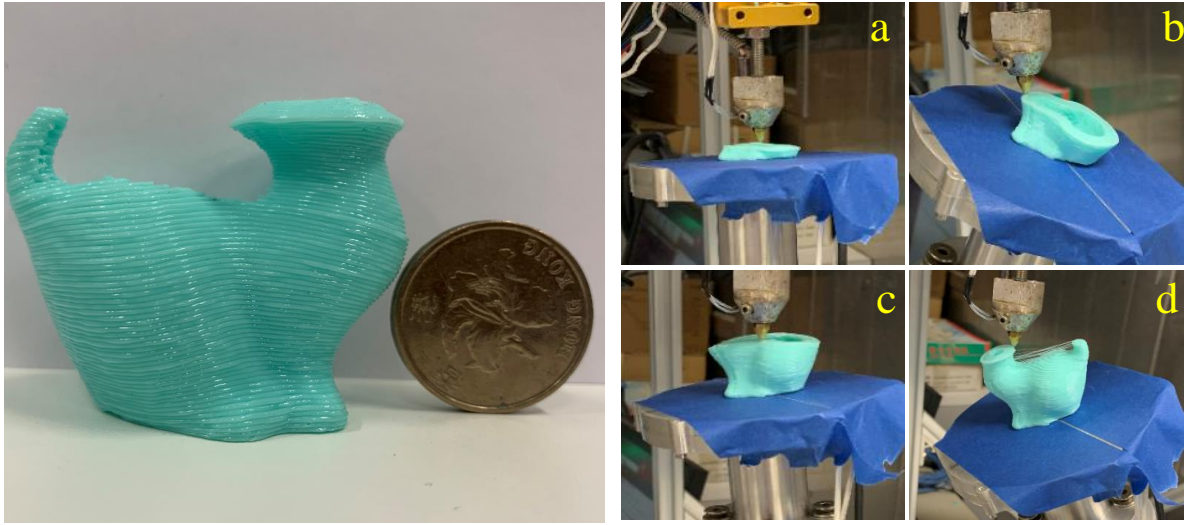

(c)

**Figure 24** The actual printing processes: (a) Y model; (b) bunny; (c) kitten

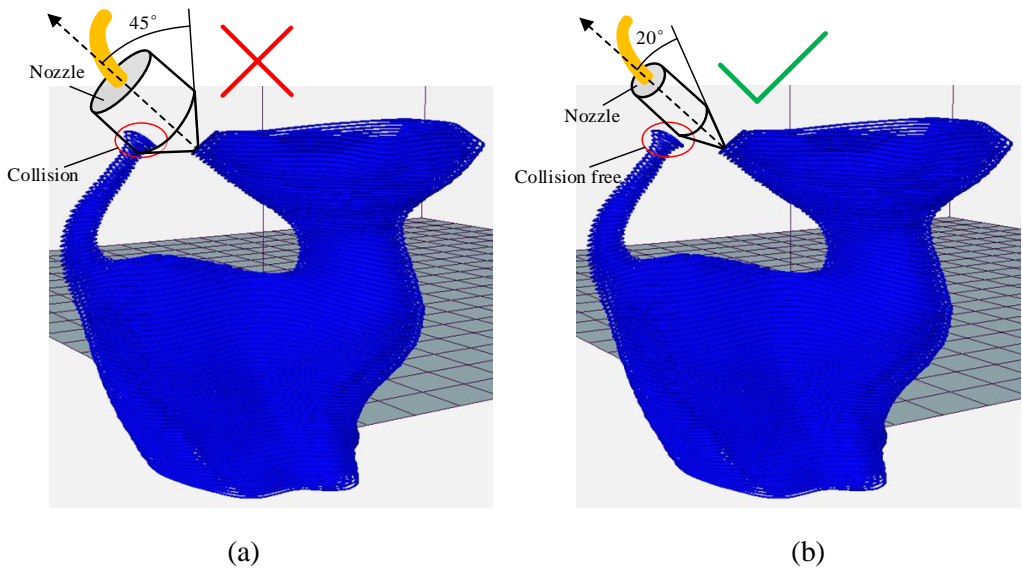

(a)　　　　　　　　　　　　　(b)

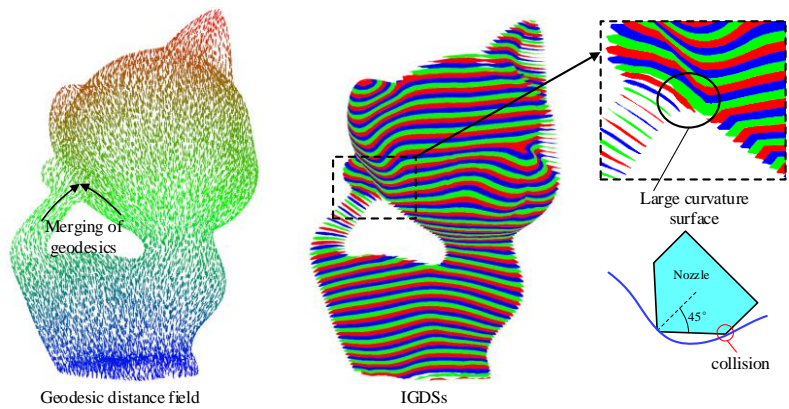

(c)

**Figure 25** The collision between the nozzle and the in-process workpiece of the kitten model: (a) collisions exist when the nozzle angle is 45°; (b) collision-free when the nozzle angle is reduced to 20° (c) large curvature IGDSs of the kitten model

Table 3 Parameters of the printing paths for the three parts

| Part | Time for printing path generation (s) | Path length (mm) | Actual printing time (min) |
|---|---|---|---|
| Y model | 74 | 73288 | 233.9 |
| Bunny | 266 | 74915 | 257.2 |
| Kitten | 217 | 89583 | / |

**4.2 Printing sequence optimization**

In this section, we test and evaluate the three different printing sequence traversal algorithms (i.e. the *LPT*, the *DPT*, and Algorithm 3 (A3)) on a tree-structured part with three branches. As shown in Figure 26, the part is decomposed into 151 layers by setting the geodesic distance interval to 0.6 mm, and the total number of the IGDSs is 311. Table 4 and Figure 27 show the simulation results of different cases. When the nozzle angle (NA) is 75°, the printing sequence generated by the *LPT* is collision-free, which requires 162 retractions and the total air-move path length is 2382 mm. However, the *DPT* fails to generate a collision-free printing sequence, although the number of retractions (only 2) and the total air-move path length (only 101 mm) would be much smaller. The A3 algorithm successfully plans a collision-free printing sequence with fewer retractions (24) and shorter path length (654 mm) as compared with those of *LPT*. Apparently, the reduced number of retractions and the air-move path length have decreased the nozzle angle (see Figure 27). Because the calculation of PCSs for each IGDS involves collision check, the time complexity of the A3 is the largest, i.e., $O(k^2*m)$, where *k* and *m* are the number of IGDSs and the average number of vertices of each IGDS, respectively. Figure 28 shows some snapshots of the actual printing processes of the A3 and *LPT* printing paths when the nozzle angle is 45°. (Note that we did not compare with the *DPT* path since it failed to print due to unresolvable collisions.) It can be found that, when compared to that of the *LPT* path, the filament drag problem was mitigated considerably by our A3 path owing to its significantly reduced retractions, which lead to much higher printing quality.

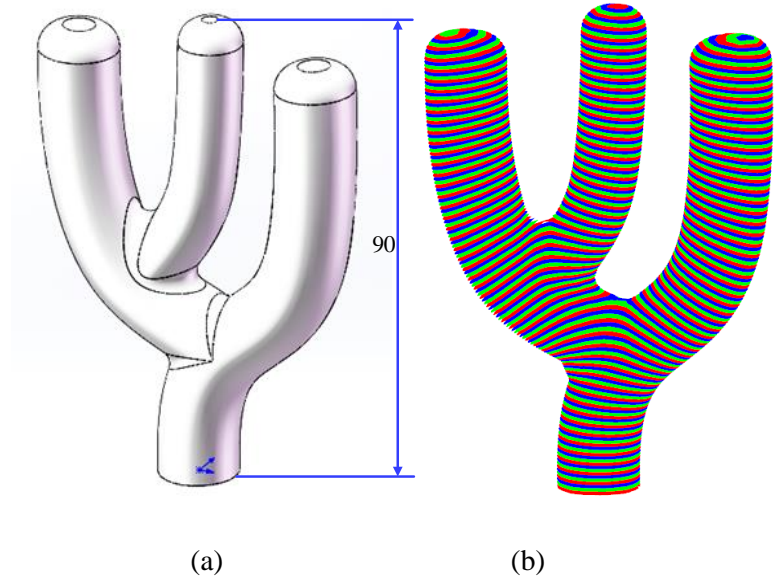

(a)            (b)

**Figure 26** A three-branch model and its IGDSs: (a) model; (b) IGDSs

Table 4 Simulation results of the three traversal algorithms

| Algorithm | Number of retractions | Air-move path length (mm) | Is collision-free or not | Running time (s) |
|---|---|---|---|---|
| LPT, NA: 75° | 162 | 2382 | Yes | 0 |
| DPT, NA: 75° | 2 | 101 | No | 0 |
| A3, NA: 75° | 24 | 654 | Yes | 271 |
| A3, NA: 60° | 17 | 371 | Yes | 177 |
| A3, NA: 45° | 7 | 227 | Yes | 277 |
| A3, NA: 30° | 5 | 146 | Yes | 154 |
| A3, NA: 15° | 4 | 125 | Yes | 155 |
| A3, NA: 1° | 3 | 108 | Yes | 242 |

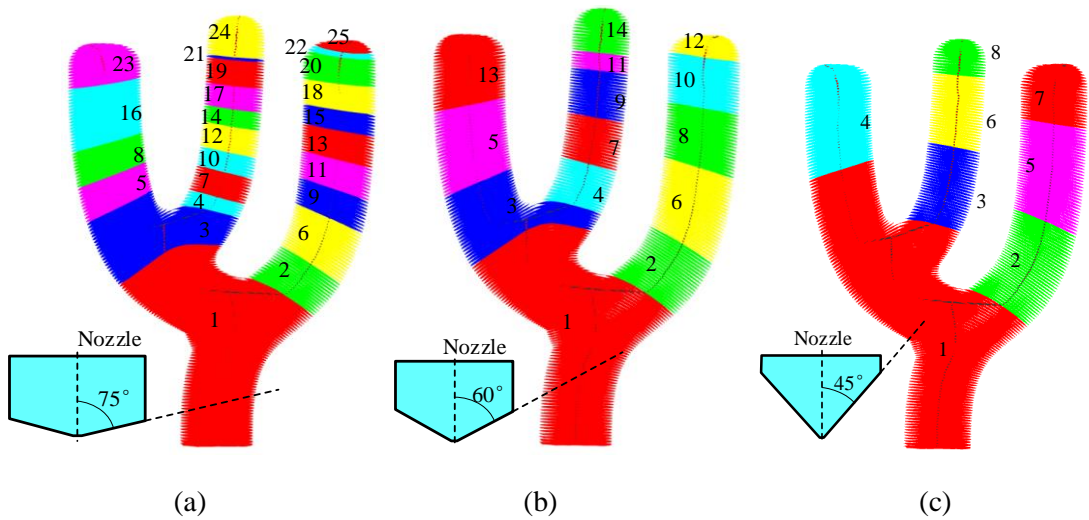

(a)            (b)            (c)

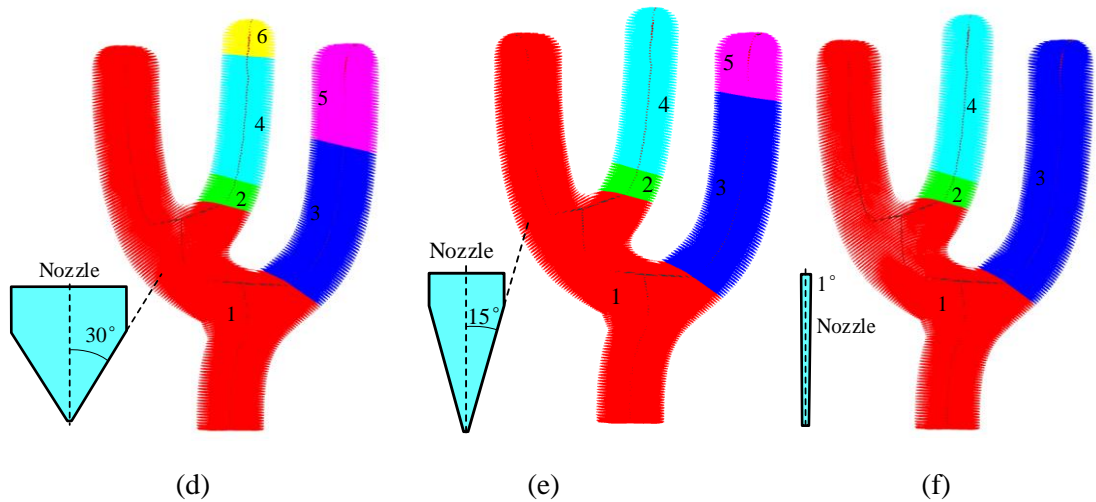

**Figure 27** Printing sequences generated by A3 under different nozzle angles: (a) 75°; (b) 60°; (c) 45°; (d) 30°; (e) 15°; (f) 1°.

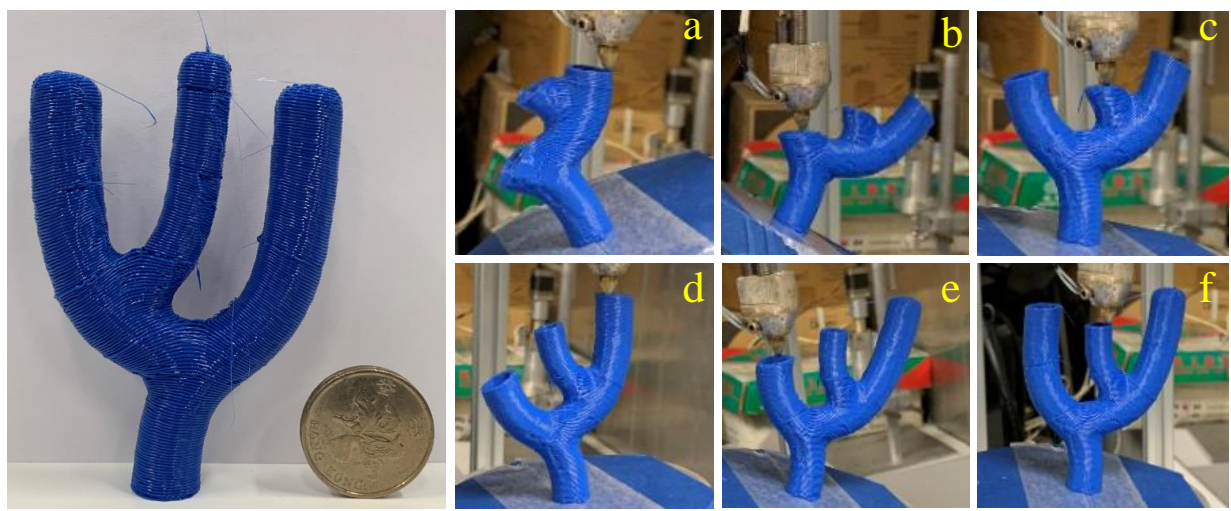

(a)

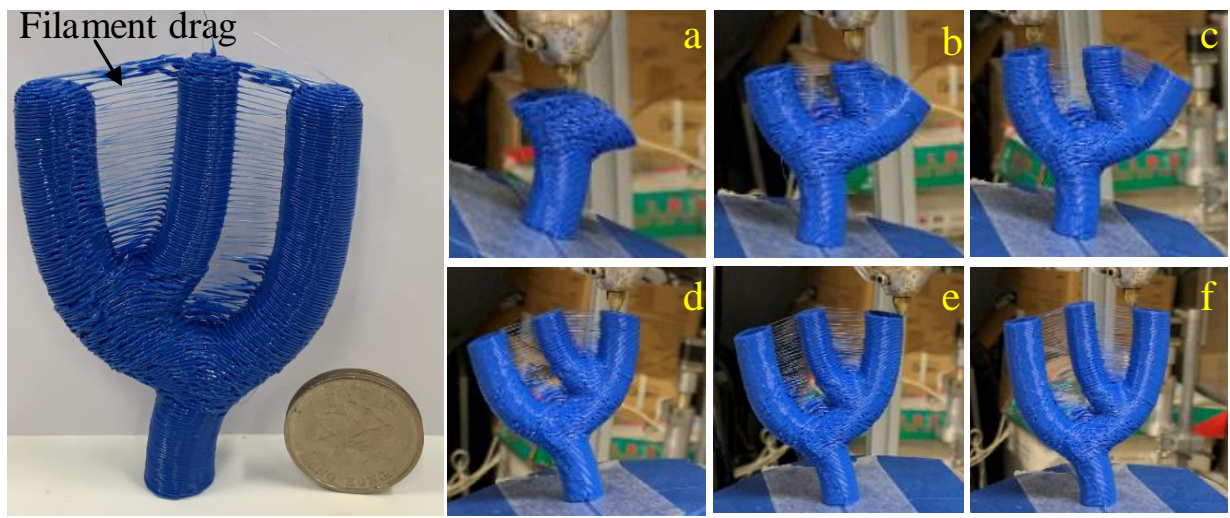

(b)

**Figure 28** Actual printing processes when the nozzle angle is 45°: (a) A3, printing path length is 39946 mm, printing time is 150 min; (b) LPT, printing path length is 42101 mm, printing time is 155 min.

## 5. Conclusion

The main objective of the work of this paper is to devise a new method for generating curved slicing layers for multi-axis support-free printing of freeform solids. Current methods are either too computationally demanding due to the voxelization or susceptible to layer-intersection problem. Inspired by the intrinsic characteristics of geodesics, we propose to first define a geodesic distance field inside the solid and then use the iso-geodesic distance surfaces (IGDSs) of this field to be the curved slicing layers. As both the proposed geodesic distance field and the IGDSs are calculated or interpolated on a tetrahedral mesh of the solid, their computation is robust, time-efficient, and simple to implement. Due to the nature of the field, the generated IGDSs are guaranteed to be clear of each other and they form a partial ordering of printing from bottom-up in a natural and smooth way, which will avoid large orientation change of the nozzle during the printing process, thereby improving the stability of the printer and saving the printing time. In addition, we present a printing sequence optimization algorithm for establishing a total ordering of the IGDSs which, while respecting the collision-free requirement, tries to minimize the air-move path length of the nozzle. The results of both computer simulation and physical printing experiments have given a positive confirmation of the proposed methods.

Regarding the future research on the subject, first, currently our printing sequencing method only supports a tree-structured part and, if the intrinsic topology of the IGDSs is a graph rather than a tree, we need first manually cut the graph into a tree and then apply the proposed printing sequencing algorithm. As there are many ways to cut a graph, this approach obviously would bring in new uncertainties and it could have many failure cases. A genuine graph traversal algorithm thus needs to be developed for a solid of genus-$n$ ($n > 0$). Secondly, it is conceivable that, even under the most conservative *LPT*, there can be cases when the collision cannot be avoided on a skeleton tree. On the other hand, as there are many ways to decompose a solid and generate curved slicing layers (e.g., [13] and [14]), a solid that fails our 3D geodesics-based volume decomposition and printing sequencing algorithm might well be printable without collision under other strategies of volume decomposition and printing sequencing. One plausible solution is that, rather than given a base, we freely select a base (including both its location and the size) on the boundary of the solid to define the geodesic distance field and the corresponding IGDSs so that their corresponding skeleton tree will be printable at least under *LPT*. Alternatively, for a given solid we could combine the proposed 3D geodesics-based volume decomposition with other types of decomposition and come up with a different set of curved slicing layers and their printing sequence that are better in dealing with the collision. All these will be our future research topics.

## 6. Acknowledgement